\documentstyle[12pt]{article}
\def\bea{\begin{eqnarray}}
\def\eea{\end{eqnarray}}
\def\nn{\nonumber}
\def\beq{\begin{equation}}
\def\eeq{\end{equation}}
\def\ba{\beq\new\begin{array}{c}}
\def\ea{\end{array}\eeq}
\def\be{\ba}
\def\ee{\ea}

\makeatletter
\newdimen\normalarrayskip              
\newdimen\minarrayskip                 
\normalarrayskip\baselineskip
\minarrayskip\jot
\newif\ifold             \oldtrue            \def\new{\oldfalse}
\def\arraymode{\ifold\relax\else\displaystyle\fi} 
\def\eqnumphantom{\phantom{(\theequation)}}     
\def\@arrayskip{\ifold\baselineskip\z@\lineskip\z@
     \else
     \baselineskip\minarrayskip\lineskip2\minarrayskip\fi}
\def\@arrayclassz{\ifcase \@lastchclass \@acolampacol \or
\@ampacol \or \or \or \@addamp \or
   \@acolampacol \or \@firstampfalse \@acol \fi
\edef\@preamble{\@preamble
  \ifcase \@chnum
     \hfil$\relax\arraymode\@sharp$\hfil
     \or $\relax\arraymode\@sharp$\hfil
     \or \hfil$\relax\arraymode\@sharp$\fi}}
\def\@array[#1]#2{\setbox\@arstrutbox=\hbox{\vrule
     height\arraystretch \ht\strutbox
     depth\arraystretch \dp\strutbox
     width\z@}\@mkpream{#2}\edef\@preamble{\halign
\noexpand\@halignto
\bgroup \tabskip\z@ \@arstrut \@preamble \tabskip\z@ \cr}%
\let\@startpbox\@@startpbox \let\@endpbox\@@endpbox
  \if #1t\vtop \else \if#1b\vbox \else \vcenter \fi\fi
  \bgroup \let\par\relax
  \let\@sharp##\let\protect\relax
  \@arrayskip\@preamble}
\def\eqnarray{\stepcounter{equation}%
              \let\@currentlabel=\theequation
              \global\@eqnswtrue
              \global\@eqcnt\z@
              \tabskip\@centering
              \let\\=\@eqncr
              $$%
 \halign to \displaywidth\bgroup
    \eqnumphantom\@eqnsel\hskip\@centering
    $\displaystyle \tabskip\z@ {##}$%
    \global\@eqcnt\@ne \hskip 2\arraycolsep
         $\displaystyle\arraymode{##}$\hfil
    \global\@eqcnt\tw@ \hskip 2\arraycolsep
         $\displaystyle\tabskip\z@{##}$\hfil
         \tabskip\@centering
    &{##}\tabskip\z@\cr}
\begingroup\ifx\undefined\newsymbol \else\def\input#1 {\endgroup}\fi
\input amssym.def \relax
\input amssym
\newfont{\hr}{msbm10}
\newfont{\ams}{msam10}

\font\numbers=cmss12
\font\upright=cmu10 scaled\magstep1
\def\stroke{\vrule height8pt width0.4pt depth-0.1pt}
\def\topfleck{\vrule height8pt width0.5pt depth-5.9pt}
\def\botfleck{\vrule height2pt width0.5pt depth0.1pt}
\def\Zmath{\vcenter{\hbox{\numbers\rlap{\rlap{Z}\kern 0.8pt\topfleck}\kern
2.2pt
                   \rlap Z\kern 6pt\botfleck\kern 1pt}}}
\def\Qmath{\vcenter{\hbox{\upright\rlap{\rlap{Q}\kern
                   3.8pt\stroke}\phantom{Q}}}}
\def\Nmath{\vcenter{\hbox{\upright\rlap{I}\kern 1.7pt N}}}
\def\Cmath{\vcenter{\hbox{\upright\rlap{\rlap{C}\kern
                   3.8pt\stroke}\phantom{C}}}}
\def\Rmath{\vcenter{\hbox{\upright\rlap{I}\kern 1.7pt R}}}
\def\Z{\ifmmode\Zmath\else$\Zmath$\fi}
\def\Q{\ifmmode\Qmath\else$\Qmath$\fi}
\def\N{\ifmmode\Nmath\else$\Nmath$\fi}
\def\C{\ifmmode\Cmath\else$\Cmath$\fi}
\def\R{\ifmmode\Rmath\else$\Rmath$\fi}

\newcounter{app}

\def\app{\setcounter{equation}{0}
\def\theequation{\Alph{app}.\arabic{equation}}\par
   \addvspace{4ex}
   \@afterindentfalse
  \secdef\@app\@dapp}

\newcommand\@app{\@startsection {app}{1}{0ex}%
                                   {-3.5ex \@plus -1ex \@minus -.2ex}%
                                   {2.3ex \@plus.2ex}%
                                   {\normalfont\Large\bf}}
\def\@dapp#1{%
{\parindent \z@ \raggedright  \bf #1}\par\nobreak}
\def\l@app#1#2{\ifnum \c@tocdepth >\z@
    \addpenalty\@secpenalty
    \addvspace{1.0em \@plus\p@}%
    \setlength\@tempdima{8em}%
    \begingroup
      \parindent \z@ \rightskip \@pnumwidth
      \parfillskip -\@pnumwidth
      \leavevmode \bfseries
      \advance\leftskip\@tempdima
      \hskip -\leftskip
      #1\nobreak\hfil \nobreak\hb@xt@\@pnumwidth{\hss #2}\par
    \endgroup\fi}
\newcounter{sapp}[app]

\def\sapp{\def\theequation{\Alph{app}.\arabic{equation}}
\par
\@afterindentfalse
  \secdef\@sapp\@dsapp}
\newcommand{\@sapp}{\@startsection{sapp}{2}{\z@}%
                                     {-3.25ex\@plus -1ex \@minus -.2ex}%
                                     {1.5ex \@plus .2ex}%
                                     {\normalfont\large\bfseries}}

\def\@dsapp#1{%
{\parindent \z@ \raggedright  \bf #1
}\par\nobreak}
\newcommand{\l@sapp}{\@dottedtocline{2}{1.5em}{2.3em}}

\def\res{{\rm res}}

\def\2{{1\over 2}}
\def\N2{${\cal N}=2$}

\def\be{ \begin{eqnarray} }
\def\ee{ \end{eqnarray} }

\def\bea{\begin{eqnarray}}
\def\eea{\end{eqnarray}}
\def\nn{\nonumber}

\def\beq{\begin{equation}}
\def\eeq{\end{equation}}
\def\ba{\beq\new\begin{array}{c}}
\def\ea{\end{array}\eeq}
\def\be{\ba}
\def\ee{\ea}

\def\theequation{\thesection.\arabic{equation}}

\begin{document}
\begin{flushright}
ITEP/TH-10/98\\
FIAN/TD-01/98\\
hepth/9802007
\end{flushright}
\vspace{0.5cm}
\begin{center}
{\LARGE \bf RG Equations from Whitham Hierarchy}
\vspace{0.5cm}

\setcounter{footnote}{1}
\def\thefootnote{\fnsymbol{footnote}}
{\Large A.Gorsky\footnote{ITEP, Moscow
117259, Russia; e-mail address: gorsky@vx.itep.ru},
A.Marshakov\footnote{Theory
Department, Lebedev Physics Institute, Moscow
~117924, Russia; e-mail address: mars@lpi.ac.ru}\footnote{ITEP,
Moscow 117259, Russia; e-mail address:
andrei@heron.itep.ru},
A.Mironov\footnote{Theory
Department, Lebedev Physics Institute, Moscow
~117924, Russia; e-mail address: mironov@lpi.ac.ru}\footnote{ITEP,
Moscow 117259, Russia; e-mail address:
mironov@itep.ru}, A.Morozov\footnote{ITEP, Moscow
117259, Russia; e-mail address: morozov@vx.itep.ru}
}\\
\end{center}
\bigskip
\begin{quotation}
The second derivatives of prepotential with respect to Whitham time-variables
in the Seiberg-Witten theory are expressed in terms of Riemann
theta-functions.
These formulas give a direct transcendental generalization of
algebraic ones for the Kontsevich matrix model.
In particular case they provide an explicit derivation
of the renormalization group (RG) equation
proposed recently in papers on the Donaldson theory.
\end{quotation}
\setcounter{footnote}{0}
\section{Introduction}
\setcounter{equation}{0}

The Seiberg-Witten (SW) theory \cite{SW}  provides an anzatz
for the low-energy effective action of $3d$, $4d$ and $5d$
$N=2$ SUSY Yang-Mills models.
It describes the dependence of the BPS spectra
$a^i(h_k,\Lambda)$
and effective coupling constants ${\cal T}_{ij}(h_k,\Lambda )$
on bare coupling (parameter
like $\Lambda = \Lambda_{QCD}$) and the choice
of vacua, parametrized by the vacuum expectation values
(v.e.v.) $h_k = \frac{1}{k} \langle tr\ \phi^k \rangle$
of the scalar field $\phi$ in the adjoint representation of
the gauge group $G$.
As explained in ref.\cite{GKMMM}, this essentially RG
problem has a natural interpretation in terms
of the quasiclassical (Whitham) integrable hierarchies
(see \cite{Car} for the most recent review and the list of references).
One expects that (the logarithm of)
the generating function of the topological correlators,
\be
\log Z(a,t) = \ \left\langle \exp \left( \sum_{n=2}^\infty t_n
{\rm Tr}\ \Phi^{n}\right) \right\rangle_{vac\{a\}}
\label{z}
\ee
($\Phi(\vartheta)$ is the {\it chiral} $N=2$ superfield,
$\phi$ being its lowest component), is expressed through
the prepotential (quasiclassical
$\tau$-function) ${\cal F}(\alpha,T)$ that is uniquely
determined by a family of the spectral curves,
parametrized by the "flat" moduli $\alpha^i$.
(The Seiberg-Witten prepotential ${\cal F}_{SW}(a)$ itself
is obtained when the Whitham times $T_n=\delta_{n,1}$,
then also $\alpha^i = a^i$.)
So far such relation has been established \cite{t+T}
only in the case of Generalized Kontsevich Model (GKM) \cite{GKM,UFN},
which is an example of $0+0$-dimensional quantum field theory.
Since \cite{GKMMM} the Seiberg-Witten theory belongs to the universality
class of $0+1$-dimensional integrable systems, it is a natural
affine ("1-loop") generalization of Kontsevich theory,
which can be hopefully understood in the nearest future.
The present paper contains some preliminary investigation
of this problem through the study of basic relevant
properties of the Seiberg-Witten prepotential.
This is a necessary but only the first step towards
generalization of the results of ref.\cite{t+T} from the Kontsevich
to Donaldson theory.

The prepotential theory \cite{prep}
defines ${\cal F}(\alpha,T)$ as a generating function of meromorphic
1-forms on the spectral curves and admits straightforward
evaluation of quantities like:\footnote{
We define the symbols $\oint$ and $res$ with
additional factors $(2\pi i)^{-1}$ so that
$$
{\res}_0 \frac{d\xi}{\xi}
= -{\res}_\infty \frac{d\xi}{\xi} =
\oint \frac{d\xi}{\xi} = 1
$$
This explains the appearance of $2\pi i$ factors in the
Riemann identities like (\ref{pro}) and thus in all the
definitions of section \ref{prepth} below.
Accordingly, the theta-functions are periodic
with period $2\pi i$, not unity (cf. \cite{BE}), and
$$
\frac{\partial\theta(\vec\xi|{\cal T})}{\partial{\cal T}_{ij}}
= i\pi\partial^2_{ij}\theta(\vec\xi|{\cal T})
$$
since periods of
the Jacobi transformation $\xi_i \equiv \int^\xi d\omega_i$
belong to $2\pi i \left( \Z + {\cal T}\Z\right)$.
}
\be
\frac{\partial{\cal F}}{\partial T^n} = \frac{\beta}{2\pi i n}
\sum_m  mT_m {\cal H}_{m+1,n+1}
= \frac{\beta}{2\pi in}T_1{\cal H}_{n+1} + O(T_2,T_3,\ldots)
\label{1der}
\ee
\be
\frac{\partial^2{\cal F}}{\partial \alpha^i\partial T^n}
= \frac{\beta}{2\pi in} \frac{\partial {\cal H}_{n+1}}{\partial a^i}
\label{11der}
\ee
\be
\frac{\partial^2{\cal F}}{\partial T^m\partial T^n}
= -\frac{\beta}{2\pi i} \left({\cal H}_{m+1,n+1}
+ \frac{\beta}{mn}\frac{\partial {\cal H}_{m+1}}{\partial a^i}
\frac{\partial {\cal H}_{n+1}}{\partial a^j}
\partial^2_{ij} \log \theta_E(\vec 0|{\cal T})\right)
\label{2der}
\ee
etc. In these formulas the gauge group is $G = SU(N)$,
parameter $\beta = 2N$, $m,n = 1,\ldots,N-1$,
and $T$-derivatives are taken at constant $\alpha^i$.
${\cal H}_{m,n}$ are certain homogeneous combinations of
$h_k$, defined in terms of the $h$-dependent polynomial
$P(\lambda)$ which is used to describe the Seiberg-Witten
(Toda-chain) spectral curves:
\be
{\cal H}_{m+1,n+1} =
-\frac{N}{mn}
{\res}_\infty\left(P^{n/N}(\lambda)d P^{m/N}_+(\lambda)\right)
= {\cal H}_{n+1,m+1}
\ee
and
\be
{\cal H}_{n+1} \equiv {\cal H}_{n+1,2}
= -\frac{N}{n}{\res}_\infty P^{n/N}(\lambda)
d\lambda  = h_{n+1} + O(h^2)
\ee

The SW theory itself allows one to evaluate derivatives (\ref{1der}),
(\ref{2der})
at $n=1$, since after appropriate rescaling
\be
h_k \rightarrow T_1^k h_k,\ \ \
{\cal H}_k \rightarrow T_1^k {\cal H}_k
\label{hH}
\ee
$T_1$ can be identified with $\Lambda$, and $\Lambda$  is
explicitly present in the SW anzatz. Eqs. (\ref{1der}),
(\ref{2der}) at $n=1$ are
naturally interpreted in terms of the stress-tensor anomaly,
\be
\ldots +\vartheta^4\Theta^\mu_\mu =
\beta {\rm tr} \Phi^2  = \ldots + \vartheta^4 \beta {\rm tr}
\left(G_{\mu\nu}G^{\mu\nu} + iG_{\mu\nu}\tilde G^{\mu\nu}\right),
\ee
since for any operator ${\cal O}$
\be
\frac{\partial}{\partial \log\Lambda} \langle {\cal O}\rangle \ =
\ \langle \beta {\rm Tr} \Phi^2, {\cal O}\rangle
\label{an}
\ee
Analogous interpretation for $n\geq 2$ involves anomalies of
$W_{n+1}$-structures.

For $n=1$, eq.(\ref{1der}), i.e. eq.(\ref{an}) for
${\cal O}=I$, has been derived
in \cite{Ma,Ma_list} in the form
\be
\frac{\partial{\cal F}_{SW}}{\partial\log\Lambda} =
\frac{\beta}{2\pi i} (T_1^2h_2)
\ee
Eq.(\ref{an}) for ${\cal O} = h_m$ -- the analogue of
for $n=1$ and any $m = 1,\ldots,N-1$, --
\be
\frac{\partial h_m}{\partial\log\Lambda} =
-\beta\frac{\partial h_{2}}{\partial a^i}
\frac{\partial h_{m}}{\partial a^j}
\partial^2_{ij} \log \theta_E({\vec 0}|{\cal T})
\label{lns}
\ee
Formulas of such a type first appeared for the $SU(2)$ case in \cite{MW}, and
literally in this form for the $SU(2)$ case in \cite{LNS,MM} and for the
general $SU(N)$ case in \cite{LNS}. It was deduced from its modular
properties and also from
sophisticated reasoning in the framework of the Donaldson theory
(the topological correlator (\ref{z}) is an object
from the Donaldson theory, i.e. topological subsector of
the $N=2$ SUSY Yang-Mills theory, which can be investigated with
the help of the twisting procedure).
In (\ref{lns}), one can substitute $h_m$ by any homogeneous function
of $h$'s, e.g. by ${\cal H}_m$.

Eq. (\ref{2der}) can be also rewritten as:
\be
\frac{\partial^2}{\partial T^m\partial T^n}
\left({\cal F}(\alpha,T) -
\frac{\beta^2}{4\pi i}{\cal F}^{GKM}(\alpha,T)\right)
= -\frac{\beta^2}{2\pi imn}
\frac{\partial {\cal H}_{m+1}}{\partial a^i}
\frac{\partial {\cal H}_{n+1}}{\partial a^j}
\partial^2_{ij} \log \theta_E(\vec 0|{\cal T})
\ee
where the prepotential of the Generalized Kontsevich Model \cite{t+T},
\be
{\cal F}^{GKM}(\alpha|T) \equiv \frac{1}{2N} \sum_{m,n} T_mT_n
{\cal H}_{m+1,n+1}
\ee
implicitly depends on the moduli $\alpha^i$ through
the coefficients of the polynomial $P(\lambda)$.

A direct purpose of this paper is to prove eqs.(\ref{1der}),
(\ref{2der})
for all $m,n = 1,\ldots, N-1$ by methods developed in refs.\cite{prep}.
An explicit derivation of (\ref{lns}) will be given
directly in terms of calculus on Riemann surfaces \cite{Fay,calc},
without appealing to $4d$ topological theory  (see also \cite{AMZ} for
related consideration).

Sections \ref{SWA} and \ref{prepth} contain a brief summary of
the Seiberg-Witten anzatz and prepotential theory (restricted
to one complex dimension -- to spectral {\it curves}).
In sections \ref{specifics} and \ref{evaluation}
it is further specialized for the particular case
of SW (Toda-chain) curves, explicit expressions are given
for the relevant meromorphic forms and the formulas
(\ref{1der})-(\ref{2der}) are explicitly derived.
An independent short proof of relation (\ref{lns}) is given
in s.\ref{dirder}, while the derivation of (\ref{lns}) directly
from (\ref{1der}) and (\ref{2der}) is presented in s.\ref{cocheck}.
Remaining problems are summarized in Conclusion.
Appendix $A$ (an extract from \cite{Fay}) contains a proof
of the relation connecting two canonical bi-differentials on Riemann
surfaces, which play the central role in reasoning
of ss.\ref{evaluation} and \ref{dirder}. Appendix $B$ is devoted
to the explicit check of eq.(\ref{lns}) in the simplest case of genus one
curves (i.e. for the $SU(2)$ gauge group).

\section{The SW Anzatz \label{SWA}}
\setcounter{equation}{0}

The simplest description of the SW anzatz \cite{SW}
exists in terms of finite-dimensional integrable systems
or 0+1 dimensional integrable field
theory \cite{GKMMM,int,SCh,int_rev,br,WDVV}.  With any $N=2$ SUSY gauge
theory one associates an integrable system with the Lax 1-form $L(\xi)d\xi$
-- a section of a holomorphic bundle over a bare spectral curve.  Its
eigenvalue, $dS_{SW}(\xi)$ is a meromorphic 1-form on the {\it full} spectral
curve.  The periods
\be a^i = \oint_{A_i} dS_{SW}, \ \ \ a^D_i = \oint_{B^i}
dS_{SW}
\label{permin}
\ee
depend on the Hamiltonians
$h_k = \frac{1}{k}tr L^k$ of integrable system and
a characteristic feature of the Lax operator is that their
variations w.r.t. moduli,
\be
\frac{\partial dS_{SW}}{\partial u_k} = dv^k,
\label{derdS}
\ee
are {\it holomorphic} differentials on the full
spectral curve so that ${\cal T}_{ij} = \partial a^D_i/\partial a^j$
is the {\it symmetric} period matrix and,
thus, $a^D_i$ can be represented as $a^i$-derivatives,
\be a^D_i = \frac{\partial{\cal F}_{SW}}{\partial a^i}
\label{prepmin}
\ee
of the {\it prepotential} ${\cal F}_{SW}(a)$.

The variation of $dS_{SW}$ w.r.t. moduli depends on the choice
of coordinate on the curve, which is fixed under the
variation.
The difference is always a total derivative, but of
a function which is multivalued or has singularities. The total-derivatives
of a single-valued function do not change $a^i$ and $a^D_i$ --
the ordinary $A$- and $B$-periods of $dS_{SW}$.
However, the Whitham dynamics depends on more delicate
characteristics such as coefficients of expansion
near the singularities, which are affected by total derivatives.
In this paper, we request (\ref{derdS}) to be an {\it exact} equality
(not modulo total derivatives).
This fixes a distinguished parametrization (coordinate)
of the full spectral curve.

In the framework of the $N=2$ supersymmetric Yang-Mills theory,
the Hamiltonians $h_k$ are associated with the v.e.v.
$h_k = \frac{1}{k}\langle tr \phi^k \rangle$ which spontaneously break
the gauge symmetry down to abelian one, while ${\cal F}_{SW}(a^i)$
is the prepotential of effective abelian $N=2$ SUSY theory at
low energies, $a^i$ being the lowest components of abelian
$N=2$ superfield.

In present paper, we concentrate on the case of pure
gauge $N=2$ SUSY Yang-Mills theory in $4$ dimensions
with the gauge group $SU(N)$.
Then the relevant $(0+1)d$ integrable system is periodic Toda chain
of the length $N$,
the bare spectral curve is double-punctured sphere,
$d\xi = \frac{dw}{w}$, $L(w)$ is $N\times N$ matrix,
the full spectral curve is given by
\be
\det_{N\times N} \left( L(w) - \lambda\right) = 0,
\ee
i.e.
\be
P(\lambda) = \Lambda^{N}\left(w + \frac{1}{w}\right),
\label{cur}
\ee
where
\be
P(\lambda) = \lambda^{N} - \sum_{k=2}^{N}
u_k\lambda^{N-k} = \prod_{i=1}^{N}(\lambda - \lambda_i)
\ee
and
\be
u_k = (-)^{k+1}\sum_{i_1<\ldots <i_k}\lambda_{i_1}\ldots\lambda_{i_k}
\ee
are the Schur polynomials of
$h_k = \frac{1}{k}\sum_{i=1}^{N} \lambda_i^k$:
\be
\log \left( \lambda^{-N}P(\lambda)\right) =
-\sum_k \frac{h_k}{\lambda^k}
\label{Schur}
\ee
Here $u_0 = 1,\ u_1 = 0, u_2 = h_2, u_3 = h_3,
u_4 = h_4 - \frac{1}{2}h_2^2, \ldots$

Alternative representation of the same family of spectral curves
is in the hyperelliptic form:
\be
Y^2 = P^2(\lambda) - 4\Lambda^{2N}, \ \ \
Y = \Lambda^{N}\left(w - \frac{1}{w}\right)
\label{hell}
\ee
The curves of this family
are simultaneously $2$-fold and $N$-fold coverings of
punctured Riemann spheres parametrized by $\lambda$
and $w$ respectively.
Among all hyperelliptic curves, the Toda-chain
ones (\ref{hell}) are distinguished by the property that
there exists a function $w = \frac{1}{2}\Lambda^{-N}(P + Y)$ which
has $N$-fold pole and $N$-fold zero (since
$w^{-1} = \frac{1}{2}\Lambda^{-N}(P - Y)$)
at the "infinities" $\lambda = \infty_\pm$
on two sheets of the hyperelliptic curve.

From (\ref{cur}) and (\ref{hell}), we get generic
relation between variations of different parameters:
\be
\delta P + P'\delta\lambda = NP\delta\log\Lambda +
Y\frac{\delta w}{w}
\label{varhell}
\ee
with $\delta P = -\sum \lambda^{N-k}\delta u_k$
and $P' = \partial P/\partial\lambda$
which can be used in evaluation of various derivatives.
In particular, on a given curve (i.e. for fixed $u_k$
and $\Lambda$),
\be
\frac{dw}{w} = \frac{P'd\lambda}{Y}
\ee
\be
dS_{SW} = \lambda \frac{dw}{w} = \frac{\lambda dP}{Y}
\label{dssw}
\ee
and
\be
\left.\frac{\partial dS_{SW}}{\partial u_k}\right|_{w = const} =
\frac{\lambda^{N-k}}{P'}\frac{dw}{w} =
\frac{\lambda^{N-k}d\lambda}{Y} = dv^k, \ \ \
k = 2,\ldots,N
\label{dv}
\ee
are $g = N-1$ holomorphic 1-forms on the curve (\ref{hell})
of genus $g=N-1$. Their $A$-periods
\be
\sigma^{ik} = \oint_{A_i} dv^k = \frac{\partial a^i}{\partial u_k}
\label{sigma}
\ee
and $d\omega_i = \sigma^{-1}_{ik}dv^k$ are the {\it canonical}
holomorphic 1-differentials,
\be
\oint_{A_i} d\omega_j = \delta^i_j, \ \ \
\oint_{B^i} d\omega_j = {\cal T}_{ij}
\label{candif}
\ee
where ${\cal T}_{ij}$ is the period matrix of (\ref{hell}).
In (\ref{dv}),
we assumed that coordinate $w$ is constant when moduli
$u_k$ are varied. It is necessary to satisfy (\ref{derdS}).
If, instead, $\lambda$ is constant, one would get
\be
\left.\frac{\partial dS_{SW}}{\partial u_k}\right|_{\lambda = const} =
\frac{\lambda^{N-k}d\lambda}{Y}
- d\left(\frac{\lambda^{N-k+1}}{Y}\right)
\ee
and the total derivative in the r.h.s. would prevent one from direct
applying of the reasoning of next sections
\ref{prepth} and \ref{specifics}.
In what follows we assume that all moduli-derivatives are
taken at constant $w$, without mentioning this explicitly.

The periods
\be
a^i = \oint_{A_i} dS_{SW}
\label{aper}
\ee
define $a^i$ as functions of $u_k$ (i.e. $h_k$) and
$\Lambda$, or, inverse, $u_k$ as functions of $a^i$
and $\Lambda$. In this paper, we are going to evaluate,
among other things,
$\partial u_k/\partial\log\Lambda |_{a^i = const}$.
From (\ref{aper}),
\be
\sum_k \left(\left.
\frac{\partial u_k}{\partial\log\Lambda}\right|_{a_i=const}\right)
\oint_{A_i} \frac{\partial dS_{SW}}{\partial u_k} +
\oint_{A_i}\frac{\partial dS_{SW}}{\partial\log\Lambda} = 0
\ee
so that
\be
\sum_k \frac{\partial u_k}{\partial\log\Lambda}
\frac{\partial a^i}{\partial u_k} =
-\oint_{A_i}\frac{\partial dS_{SW}}{\partial\log\Lambda} =
-N\oint_{A_i} \frac{P}{P'}\frac{dw}{w} =
-N\oint_{A_i} \frac{Pd\lambda}{Y} = \nn \\ =
-N\oint_{A_i}\frac{(P+Y)d\lambda}{Y} =
-2N\Lambda^{N}\oint_{A_i}\frac{wd\lambda}{Y} =
-2N\Lambda^{N}\oint_{A_i} w dv^{N}
\label{dudll}
\ee
We use this relation in explicit derivation
of eq.(\ref{lns}) in s.\ref{dirder} below.

Another, much simpler relation can be deduced right now.
Since
\be
NPd\lambda = \lambda dP - \sum_k ku_k\lambda^{N-k}d\lambda
\label{polrel}
\ee
and due to (\ref{dssw})-(\ref{sigma}), one obtains from
(\ref{dudll}):
\be
-\sum_k \frac{\partial u_k}{\partial\log\Lambda}
\frac{\partial a^i}{\partial u_k} =
a^i - \sum_k ku_k\frac{\partial a^i}{\partial u_k}
\ee
or
\be
\frac{\partial u_k}{\partial\log\Lambda}
= ku_k - a^i\frac{\partial u_k}{\partial a^i}
\label{dimrel}
\ee
This, of course, follows from dimensional considerations.

Below we put $\Lambda = 1$ to simplify formulas
(one can also say that $\Lambda$ is absorbed in rescaling
of $\lambda$, $u$, $h$ and $T$'s).

\section{Prepotential theory \label{prepth}}
\setcounter{equation}{0}

The prepotential theory, as developed in \cite{prep},
defines the prepotential (quasiclassical or Whitham $\tau$-function)
for any finite-gap solution of the KP/Toda hierarchy, i.e. for the following
set of data:

-- complex curve (Riemann surface) of genus $g$;

-- a set of punctures (marked points) on it;

-- coordinates $\xi$ in the vicinities of the punctures.

In this section we consider the simplest case of a single
puncture, say at $\xi = \xi_0 = 0$.

Given this set of data, one can introduce meromorphic
1-differentials with the poles at $\xi=\xi_0=0$ only such that
\be
d\Omega_n = \left( \xi^{-n-1} + O(1) \right)d\xi,
\ \ \  n\geq 1
\label{canbas}
\ee
This condition defines $d\Omega_n$ up to arbitrary
linear combination of $g$ holomorphic differentials
$d\omega_i$, $i = 1,\ldots,g$ and there are two different ways to fix this
ambiguity.

The first way is to require that $d\Omega_n$ have vanishing
$A$-periods,
\be
\oint_{A_i} d\Omega_n = 0    \ \ \ \forall i,n
\label{vanAp}
\ee
We keep the notation $d\Omega_n$ for {\it such} differentials.
Their generating functional,
\be
W(\xi,\zeta) = \sum_{n=1}^\infty n\zeta^{n-1} d\zeta d\Omega_n(\xi)
\label{Wexp}
\ee
is well known in the theory of Riemann surfaces.
It can be expressed through the Prime form $E(\xi,\zeta) =
\frac{\theta_*(\vec\xi - \vec\zeta)}
{\nu_*(\xi)\nu_*(\zeta)}$:\footnote{
For example, for genus $g=1$
$$
d\Omega_1 = (\wp (\xi) - const)d\xi, \ d\Omega_2 = -\frac{1}{2}\wp'(\xi)d\xi,
\ \ldots, d\Omega_n = \frac{(-)^{n+1}}{n!}\partial^{n-1}\wp(\xi)d\xi
$$
and
$$
W(\xi,\zeta) = \sum_{n=1}^\infty \frac{(-)^{n+1}\zeta^{n-1}}{(n-1)!}
\frac{\partial^{n-1}}{\partial\xi^{n-1}}\wp(\xi)d\xi d\zeta
- const\cdot d\xi d\zeta =
\wp(\xi - \zeta)d\xi d\zeta =
\partial_\xi \partial_\zeta \log \theta_*(\xi - \zeta)
$$
where $*$ denotes the odd theta-characteristic. For $g=1$ $\nu_*^2(\xi)
=\theta_{*,i}(0)d\omega_i$ is just $d\xi$.
}
\be
W(\xi,\zeta) = \partial_\xi \partial_\zeta \log E(\xi,\zeta)
\label{WE}
\ee
One needs to take care about the choice of coordinate $\zeta$
in order to guarantee that $d\Omega_n$ in this expansion indeed satisfies
the condition (\ref{canbas}). We shall check this property explicitly when
(\ref{Wexp}) will be used in s.\ref{specifics} below.
Such  $W(\xi,\zeta)$ has a second order pole on diagonal,
\be
W(\xi,\zeta) \sim \frac{d\xi d\zeta}{(\xi - \zeta)^2} + O(1) =
\sum_{n=1}^\infty n\frac{d\xi}{\xi^{n+1}}\zeta^{n-1}d\zeta + O(1)
\ee
and this explains the choice of expansion coefficients in (\ref{Wexp}).

The second way is to impose the condition like
(\ref{derdS}),
\be
\frac{\partial d\hat\Omega_n}{\partial\ {\rm moduli}} =
{\rm holomorphic}
\label{derdS1}
\ee
We shall concentrate on the situation when the number of moduli
is equal to the genus of the spectral curve, which is the case
in the SW theory. This time we introduce a
different generating functional for $d\hat\Omega_n$
with infinitely many auxiliary parameters $T_n$:
\be
dS = \sum_{n\geq 1} T_n d\hat\Omega_n =
\sum_{i=1}^g \alpha^i d\omega_i + \sum_{n\geq 1} T_nd\Omega_n
\label{dSexp}
\ee
(the generating functionals like (\ref{Wexp}) with a single
expansion parameter $\zeta$ are obtained
as a variation of this one under Miwa-like transform,
$\Delta T_n \sim \zeta^n$).
The periods
\be
\alpha^i = \oint_{A_i} dS
\label{perdef}
\ee
can be considered as particular coordinates on the moduli space.
Note that these periods are not just the same as (\ref{permin}),
eq.(\ref{perdef}) defines $\alpha^i$ as functions of $h_k$ and $T^n$,
or, alternatively, $h_k$ as functions of $\alpha^i$ and $T^n$
so that derivatives $\partial h_k/\partial T^n$ are non-trivial.
In what follows we shall consider $\alpha^i$ and $T^n$,
\be
T_n = {\res}_{\xi=0}\ \xi^n dS(\xi)
\label{Tdef}
\ee
as independent variables so that partial derivatives w.r.t.
$\alpha^i$ are taken at constant $T^n$ and partial derivatives
w.r.t. $T^n$ are taken at constant $\alpha^i$. In particular,
\be
\frac{\partial dS}{\partial \alpha^i} = d\omega_i, \ \ \
\frac{\partial dS}{\partial T^n} = d\Omega_n
\label{derfdS}
\ee

Now one can introduce the {\it prepotential}
${\cal F}(\alpha^i,T^n)$ by an analog of conditions
(\ref{prepmin}):
\be
\frac{\partial {\cal F}}{\partial \alpha^i} = \oint_{B^i} dS, \ \ \
\frac{\partial {\cal F}}{\partial T^n} = \frac{1}{2\pi in}
 {\res}_0\ \xi^{-n}dS
\label{prepdef}
\ee
Their consistency follows from (\ref{derfdS}) and Riemann
identities. In particular,
\be
\frac{\partial^2{\cal F}}{\partial T^m\partial T^n} =
\frac{1}{2\pi in} {\res}_0\ \xi^{-n} \frac{\partial dS}{\partial T^m}
= \frac{1}{2\pi in}{\res}_0\ \xi^{-n} d\Omega_m =
\frac{1}{2\pi im} {\res}_0\ \xi^{-m} d\Omega_n
\label{secderF}
\ee
From this calculation, it is clear that the definition
(\ref{prepdef}) assumes that coordinates $\xi$ are not changed
under the variation of moduli. This means that they provide a
moduli-independent parametrization of the entire family --
like $\lambda$ or $w$ in the case of (\ref{hell}). Since
moduli-independence of $\xi$ should be also consistent with
(\ref{derdS1}), the choice of $\xi$ is strongly restricted:
to $w^{\pm 1/N}$ in the case of (\ref{hell}).

The last relation in (\ref{secderF})
(symmetricity) is just an example of the Riemann relations and it
is proved by the standard reasoning:
\be
0 = \int d\Omega_m \wedge d\Omega_n =
\oint_{A_i} d\Omega_m \oint_{B^i} d\Omega_n -
\oint_{A_i} d\Omega_n \oint_{B^i} d\Omega_m  +
\frac{1}{2\pi i}{\res} \left(d\Omega_m d^{-1} d\Omega_n\right) =\\=
0 + 0 +  \frac{1}{2\pi in}{\res}_0\ \xi^{-n} d\Omega_m -
\frac{1}{2\pi im}{\res}_0\ \xi^{-m} d\Omega_n
\label{pro}
\ee
where (\ref{vanAp}) and (\ref{canbas}) are used at the last
stage. The factors like $n^{-1}$ arise since
$\xi^{-n-1}d\xi = -d(\xi^{-n}/n)$. In next
sections \ref{specifics}, \ref{evaluation}, we shall
use a slightly different normalization $d\Omega_n \sim
w^{n/N}\frac{dw}{w} = \frac{N}{n}dw^{n/N}$, accordingly the residues
in (\ref{pro}) and (\ref{prepdef}) and (\ref{secderF})
will be multiplied by $N/n$ instead of $1/n$.

By definition, the prepotential is a homogeneous function of
its arguments $a^i$ and $T^n$ of degree 2,
\be
2{\cal F} =
\alpha^i \frac{\partial{\cal F}}{\partial \alpha^i} +
T^n\frac{\partial{\cal F}}{\partial T^n} =
\alpha^i\alpha^j
\frac{\partial^2{\cal F}}{\partial \alpha^i\partial \alpha^j} +
2\alpha^i T^n
\frac{\partial^2{\cal F}}{\partial \alpha^i\partial T^n} +
T^mT^n \frac{\partial^2{\cal F}}{\partial T^m\partial T^n}
\label{hom}
\ee
Again, this condition can be
proved with the help of Riemann identities, starting from
(\ref{prepdef}), (\ref{perdef}) and (\ref{Tdef}).
At the same time, ${\cal F}$ is not just a quadratic
function of $a^i$ and $T^n$, a non-trivial dependence on
these variables arise through the dependence of $d\omega_i$
and $d\Omega_n$ on the moduli (like $u_k$ or $h_k$)
which in their turn depend on $a^i$ and $T_n$. This
dependence is described by the {\it Whitham equations}
obtained, for example, by substitution of (\ref{dSexp}) into (\ref{derfdS})
\cite{prep}:
\be
d\hat\Omega_n + T_m\frac{\partial d\hat\Omega_n}{\partial u_l}
\frac{\partial u_l}{\partial T_m} = d\Omega_n,
\ee
i.e.
\be
\left(\sum_{m,l} T_m\frac{\partial u_l}{\partial T_m}\right)
\oint_{A_i}\frac{\partial d\hat\Omega_n}{\partial u_l}
= - \oint_{A_i} d\hat\Omega_n
\ee
The integral in the l.h.s. is expressed, according to (\ref{derdS1}),
through the integrals of holomorphic 1-differentials, while
the integral in the r.h.s. --  through the periods of $dS$.
If
\be
\oint_{A_i} \frac{\partial dS}{\partial u_l} =
\oint_{A_i} dV^l = \Sigma^{il}
\ee
then
\be
T_m\frac{\partial u_k}{\partial T_m}
= \Sigma^{-1}_{ki}\alpha^i
\label{Whitheq}
\ee
For particular families of curves these formulas can be simplified further.

\section{Specialization to the Toda-chain (SW) spectral curves
\label{specifics}}
\setcounter{equation}{0}

Our goal now is to apply eqs.(\ref{prepdef}) and (\ref{secderF})
to the particular case of the spectral curves (\ref{hell}).
The first problem is that naively there are no solutions to eq.(\ref{derdS1})
because the curves (\ref{hell}) are the spectral curves
of the {\it Toda-chain} hierarchy (not of KP/KdV type).
The difference is that the adequate description in the Toda case requires
{\it two} punctures instead of one. As already mentioned in s.2 above,
for the curves (\ref{hell}) there exists a function $w$ with the $N$-degree
pole and zero at two points (to be used for the punctures)
$\lambda = \infty_\pm$, where
$\pm$ labels two sheets of the curve in the hyperelliptic representation
(\ref{hell}), $w(\lambda = \infty_+) = \infty$,
$w(\lambda = \infty_-) = 0$.
Accordingly, there are two families of the differentials $d\Omega_n$:
$d\Omega^+_n$ with poles at $\infty_+$ and
$d\Omega^-_n$ with poles at $\infty_-$.

However, there are {\it no} differentials $d\hat\Omega^\pm$,
only $d\hat\Omega_n = d\hat\Omega^+_n + d\hat\Omega^-_n$, i.e.
condition (\ref{derdS1}) requires $d\hat\Omega_n$ to have
poles at both punctures. Moreover, the coefficients in front of
$w^{n/N}$ at $\infty_+$ and $w^{-n/N}$ at $\infty_-$
(\ref{canbas}) are just the same (in Toda-hierarchy
language, this is the Toda-{\it chain} case with the same dependence upon
negative and positive times).

The differentials $d\hat\Omega_n$ for the family (\ref{hell})
have the form:
\be
d\hat\Omega_n = R_n(\lambda)\frac{dw}{w}
= P^{n/N}_+(\lambda)\frac{dw}{w}
\label{expldO}
\ee
The polynomials $R_n(\lambda)$ of degree $n$ in $\lambda$
are defined by the property that
$P'\delta R_n - R_n'\delta P$ is a polynomial of degree
less than $N-1$. Thus, $R_n(\lambda) = P^{n/N}_+(\lambda)$,
where $\left(\sum_{k=-\infty}^{+\infty} c_k\lambda^k\right)_+
= \sum_{k=0}^{+\infty} c_k\lambda^k$.
For example:
\be
R_1 = \lambda, \nn \\
R_2 = \lambda^2 - \frac{2}{N}u_2, \nn \\
R_3 = \lambda^3 - \frac{3}{N}u_2\lambda - \frac{3}{N}u_3, \nn \\
R_3 = \lambda^4 - \frac{4}{N}u_2\lambda^2 - \frac{4}{N}u_3\lambda -
\left(\frac{4}{N}u_4 + \frac{2(N-4)}{N^2}u_2^2\right), \nn \\
\ldots
\ee
These  differentials satisfy (\ref{derdS1})
provided the moduli-derivatives are
taken at constant $w$ (not $\lambda$!). Thus, the
formalism of the previous section \ref{prepth} is applicable
for $\xi = w^{\mp 1/N}$.

The Seiberg-Witten differential $dS_{SW}$ is just
$dS_{SW} = d\hat\Omega_2$, i.e.
\be
\left.dS\right|_{T_{n}=\delta_{n,1}} = dS_{SW}, \ \ \
\left.\alpha^i\right|_{T_{n}=\delta_{n,1}} = a^i, \ \ \
\left.\alpha^D_i\right|_{T_{n}=\delta_{n,1}} = a^D_i
\ee
Thus, the data one needs for the definition of prepotential can be
fixed as
follows: the punctures are at $\lambda = \infty_\pm$, the relevant
coordinates in the vicinities of these punctures are
$\xi \equiv w^{-1/N} \sim \lambda^{-1}$ at $\infty_+$ and
$\xi \equiv w^{+1/N} \sim \lambda^{-1}$ at $\infty_-$.
The parametrization in terms of $w$, however, does not allow
to use the advantages
of hyperelliptic parametrization (\ref{hell}).
Therefore, at some stages of calculations below we shall transform
the above definitions to the coordinate $\lambda$.
For this purpose, let us introduce one more notation:
$d\tilde\Omega$ for the differentials which, in contrast to
$d\Omega$, are defined by (\ref{canbas}) with $\xi = \lambda^{-1}$
(while this condition is imposed with
$\xi = w^{\mp 1/N}$ for all $d\Omega$'s).

\section{Evaluation of the prepotential derivatives
\label{evaluation}}
\setcounter{equation}{0}

Near $\lambda = \pm\infty$ the parameter $w^{\pm 1} =
\frac{1}{2}(P\pm Y) = P(\lambda)\left(1 + O(P^{-2})\right) =
P(\lambda)\left(1 + O(\lambda^{-2N})\right)$.
Thus, in the calculations involving $w^{\pm n/N}$ with
$n < 2N$, one can substitute $w^{\pm 1/N}$ by
$P(\lambda)^{1/N}$.

\subsection{First derivatives}

After all comments we made above,
(see especially the remark after eq.(\ref{pro}))
one can write now for $n<2N$:
\be
\frac{\partial{\cal F}}{\partial T^n} =
\frac{N}{2\pi in}\left(
{\res}_{\infty_+} w^{n/N} dS + {\res}_{\infty_-} w^{-n/N} dS\right)
= \nn \\ =
\frac{N}{2\pi in}\left(
{\res}_{\infty_+} w^{n/N} + {\res}_{\infty_-} w^{-n/N}\right)
\left(\sum_m T_m P^{m/N}_+(\lambda)\right) \frac{dw}{w}
= \nn \\ =
 \frac{N^2}{i\pi n^2}\sum_m T_m {\res}_\infty \left(P^{m/N}_+(\lambda)
d P^{n/N}(\lambda)\right) =
-\frac{N^2}{i\pi n^2}\sum_m T_m {\res}_\infty \left(P^{n/N}(\lambda)
d P^{m/N}_+(\lambda)\right)
\label{firder}
\ee
Substituting this expression into (\ref{hom})
and one can compare it with the prepotential of
Generalized Kontsevich Model (GKM) \cite{t+T},
\be
{\cal F}^{(N)}_{GKM}(T) =
\sum_{m,n} \frac{T_mT_n}{2mn}
{\res}_\infty \left(P^{n/N}(\lambda)
d P^{m/N}_+(\lambda)\right)
\ee
The obvious new ingredient in the SW case
is the additional transcendental piece $\frac{1}{2}a^ia_i^D$
arising due to non-trivial spectral curve.
Of course, the dependence of the polynomial $P(\lambda)$,
i.e. coefficients $u_k$ on $T_m$ is now also transcendental --
as dictated by the Whitham equations (\ref{Whitheq}). In the
GKM case, the Whitham equations can be resolved algebraically:
\be
T_n^{(GKM)} = -\frac{N}{N-n}
{\res}_\infty P(\lambda)^{1 - n/N}d\lambda
\label{KT}
\ee
These quantities appeared first \cite{KriT} in the
context of topological Landau-Ginzburg theories (see also the first two
references of \cite{prep} and \cite{t+T}), where $P(\lambda) = W'(\lambda)$
is the first derivative of the {\it superpotential}.

If evaluated at $T_n = \delta_{n,1}$, (\ref{firder}) becomes:
\be
\frac{\partial{\cal F}}{\partial T_n} = -\frac{N^2}{i\pi n^2}
{\res}_\infty P^{n/N}(\lambda) d\lambda =
\frac{N}{i\pi n}{\cal H}_{n+1}
\ee
Here
\be
{\cal H}_{n+1} \equiv -\frac{N}{n}
{\res}_\infty P^{n/N}(\lambda) d\lambda
= \sum_{k\geq 1} \frac{(-)^{k-1}}{k!}
\left(\frac{n}{N}\right)^{k-1} \sum_{i_1+\ldots +i_k = n+1}
h_{i_1}\ldots h_{i_k} = \nn \\ =
h_{n+1} - \frac{n}{2N}\sum_{i+j = n+1} h_ih_j
+ O(h^3)
\label{Hdef}
\ee
Note that these are essentially the same algebraic combinations
of $h_k$ as (\ref{KT}), but $h_k$ as functions of $T$ are,
of course, different for the GKM and SW theories.
This completes the proof of eq.(\ref{1der}).

\subsection{Mixed derivatives}

For the mixed derivatives, one obtains
\be
\frac{\partial^2{\cal F}}{\partial \alpha^i\partial T^n} =
\oint_{B_i} d\Omega_n = \frac{1}{2\pi in} {\res}_0\ \xi^{-n} d\omega_i
= \nn \\ =
\frac{N}{2\pi in} \left(
{\res}_{\infty_+} w^{n/N} d\omega_i +
{\res}_{\infty_-} w^{n/N} d\omega_i\right) =
\frac{N}{i\pi n} {\res}_\infty \ P(\lambda)^{n/N} d\omega_i
\ee
Let us introduce a special notation for the coefficients of expansion
of $P(\lambda)^{n/N}$:
\be
P(\lambda)^{n/N} = \sum_{k= -\infty}^{n} p_{nk}^{(N)}\lambda^k
\label{Pexp}
\ee
Then,
\be
{\res}_\infty \ P(\lambda)^{n/N} d\omega_i
=  \sum_{k= -\infty}^{n} p_{nk}^{(N)}{\res}_\infty \
\lambda^k d\omega_i
\label{cadires}
\ee
The canonical differential is
\be
d\omega_j(\lambda) = \sigma^{-1}_{jk}dv^k(\lambda) =
\sigma^{-1}_{jk}\frac{\lambda^{N-k}d\lambda}{Y(\lambda)} =
\sigma^{-1}_{jk}\frac{\lambda^{N-k}d\lambda}{P(\lambda)}
\left( 1 + O(\lambda^{-2N})\right) =\\=
- \sigma^{-1}_{jk}\frac{\partial\log P(\lambda)}{\partial u_k}
d\lambda \left( 1 + O(\lambda^{-2N})\right)
\label{cadires1})
\ee
From (\ref{Schur}) and
$\sigma^{-1}_{jk} = \frac{\partial u_k}{\partial a^j}$, one gets:
\be
d\omega_j(\lambda)\left( 1 + O(\lambda^{-2N})\right)  =
\sum_{n\geq 2} \sigma^{-1}_{jk} \frac{\partial h_n}{\partial u_k}
\frac{d\lambda}{\lambda^n} =
\sum_{n\geq 1} \frac{\partial h_{n+1}}{\partial a^i}
\frac{d\lambda}{\lambda^{n+1}}
\label{doexp}
\ee
so that for $k<2N$
\be
{\res}_\infty \lambda^k d\omega_i =
\frac{\partial h_{k+1}}{\partial a^i}
\label{reslo}
\ee
Note that the residue vanishes for $k<1$. Therefore,
from (\ref{cadires}) it follows for $n < 2N$:
\be
{\res}_\infty \ w^{n/N} d\omega_i =
{\res}_\infty \ P(\lambda)^{n/N} d\omega_i
=  \sum_{k= -\infty}^{n} p_{nk}^{(N)}
\frac{\partial h_{k+1}}{\partial a^i}
=  \sum_{k= 1}^{n} p_{nk}^{(N)}
\frac{\partial h_{k+1}}{\partial a^i}
\label{cadires2}
\ee
Further,
\be
\sum_{k=-\infty}^n p_{nk}^{(N)} \delta h_{k+1} =
\sum_{k=1}^\infty \oint_\infty
\frac{d\lambda}{\lambda^{k+1}} P(\lambda)^{n/N}
\delta h_{k+1} = \nn \\ =
-\oint_\infty d\lambda P(\lambda)^{n/N} \delta\log P(\lambda)
= -\frac{N}{n}\oint_\infty \delta P(\lambda)^{n/N} d\lambda =
\delta{\cal H}_{n+1}
\label{pvarh}
\ee
and, finally,
\be
{\res}_\infty \ w^{n/N} d\omega_i =
{\res}_\infty \ P(\lambda)^{n/N} d\omega_i
= \frac{\partial{\cal H}_{n+1}}{\partial a^i}
\label{cadiresfin}
\ee
Therefore,
\be
\frac{\partial^2{\cal F}}{\partial \alpha^i\partial T^n} =
\frac{N}{i\pi n}{\res}_\infty \ P(\lambda)^{n/N} d\omega_i
=  \frac{N}{i\pi n}\frac{\partial{\cal H}_{n+1}}{\partial a^i}
\ee
This completes the proof of eq.(\ref{11der}).

\subsection{Second $T$-derivatives}
\subsubsection{First step}

The expression (\ref{firder}) is not directly applicable
to the evaluation of the second derivatives of ${\cal F}$,
because $P(\lambda)$ depends on $T$ through $u_k$,
and $u(T)$ is a solution to the transcendental
Whitham equations (\ref{Whitheq}). Therefore,
we use general formulas (\ref{secderF}):
\be
\frac{\partial^2{\cal F}}{\partial T^m\partial T^n} =
\frac{1}{2\pi in} {\res}_0 \xi^{-n} d\Omega_n =
\frac{N}{2\pi in}\left(
{\res}_{\infty_+} w^{n/N} d\Omega_m
+ {\res}_{\infty_-} w^{-n/N} d\Omega_m \right)
\label{FTT1}
\ee
It is much easier to evaluate the residues involving the
differentials $d\tilde\Omega$ instead of $d\Omega$
(see the end of the previous section \ref{specifics}).
Thus, one needs
an expression for $d\Omega^\pm_n$ through $d\tilde\Omega^\pm_n$
which is easily obtained from the asymptotics at $\lambda = \infty$:
\be
d\Omega^\pm_n = \pm\left(w^{\pm n/N} + O(1)\right)\frac{dw}{w} =
\frac{N}{n} dw^{\pm n/N} + \ldots = \nn \\ =
\frac{N}{n} d P^{n/N} + \ldots =
\frac{N}{n}\sum_{k=1}^n kp_{nk}^{(N)} \lambda^{k-1}d\lambda + \ldots =
\frac{N}{n}\sum_{k=1}^n kp_{nk}^{(N)} d\tilde\Omega^\pm_k
\label{asympt}
\ee
The dots denote the non-singular terms which are unambiguously
determined by the singular ones. Therefore, to establish formula
(\ref{asympt}), it is enough to compare only the singular terms of
$d\Omega_n$ and $d\tilde\Omega_n$ (since both these sets
form complete basis of differentials with zero $A$-periods and holomorphic
outside $+\infty$ or $-\infty$) (note that only the $k>0$ terms are singular).

Substituting (\ref{asympt}) and (\ref{Pexp}) into (\ref{FTT1}),
one gets for $m,n < 2N$:
\be
\frac{\partial^2{\cal F}}{\partial T^m\partial T^n} =
-\frac{N^2}{i\pi mn}\sum_{l=1}^m l p_{ml}^{(N)}
{\res}_\infty w^{n/N} d\tilde\Omega_l \nn \\
d\tilde\Omega_l = d\tilde\Omega_l^+ + d\tilde\Omega_l^-
\label{profor}
\ee

\subsubsection{Szeg\"o kernel}

In order to evaluate the remaining residue in (\ref{profor}),
one needs a representation of $d\tilde\Omega_n$ or their generating
bi-differential $W(\xi,\zeta)$ in hyperelliptic coordinates.
This is not a trivial problem: the scalar Green function
which is the most natural quantity to be associated with
$W(\xi,\zeta)$,
\be
\langle \partial X(\xi) \partial X(\zeta) \rangle =
\partial_\xi \partial_\zeta \log \left(
|E(\xi,\zeta)|^2\exp\left({\rm Im}(\vec\xi - \vec\zeta)
\frac{1/(4\pi i)
}{{\rm Im}{\cal T}}{\rm Im}(\vec\xi - \vec\zeta)\right)\right)
= \nn \\ =
W(\xi,\zeta)-
{1\over 2\pi i}d\omega_i(\xi)({\rm Im}{\cal T})^{-1}_{ij}d\omega_j(\zeta)
\ee
in the hyperelliptic parametrization looks as follows \cite{hegf}:
\be
\langle \partial X(\lambda) \partial X(\mu) \rangle =
\left(\int\left|\prod_{i=1}^g\frac{dv_i}{Y(v_i)}\prod_{i<j}(v_i-v_j)
\right|^2\right)^{-1}\times \nn \\ \times
\int \left\{ \left|\prod_{i=1}^g\frac{dv_i}{Y(v_i)}\prod_{i<j}(v_i-v_j)
\right|^2 \frac{1}{Y(\mu)}\frac{\partial}{\partial \lambda}
\frac{1}{\lambda - \mu}\prod_{i=1}^g\frac{\mu - v_i}{\lambda - v_i}
\left( Y(\lambda)
\phantom{\prod^a_b}  + \right.\right. \nn \\ \left.\left. +
Y(\mu)\prod_{i=1}^g\frac{\lambda - v_i}{\mu - v_i} +
\sum_{j=1}^g Y(v_j)\frac{\lambda - \mu}{v_j - \mu}
\prod_{i\neq j}\frac{\lambda - v_i}{v_j - v_i}
\right)\right\} + (\lambda \leftrightarrow \mu)
\label{GF}
\ee
This expression cannot be used effectively, but,
fortunately, a simple trick allows one to proceed
without using (\ref{GF}).

Indeed, there is another bi-differential, which has a very
simple hyperelliptic representation. It is the square of the
Szeg\"o kernel -- the fermionic Green function:
\be
\Psi_e(\xi,\zeta) = \langle \psi(\xi) \tilde\psi(\zeta) \rangle =
\frac{\theta_e(\vec\xi - \vec\zeta)}{\theta_e(\vec 0)E(\xi,\zeta)}
\ee
which is a $1/2$-differential in both variables and depends on the choice of
theta-characteristic $e$ (boundary condition for the fermions). A simple
hyperelliptic representation exists for the even non-singular half-integer
characteristics. Such characteristics are in one-to-one correspondence with
the partitions of the set of all the $2g+2$ ramification points into two
equal subsets, $\{r^+_\alpha\}$ and $\{r^-_\alpha\}$, $\alpha =
1,\ldots,g+1$, $Y(\lambda) = \prod_{\alpha =1}^{g+1} (\lambda - r^+_\alpha)
(\lambda - r^-_\alpha)$. Given these two sets, one can define:
\be
U_e(\lambda) = \sqrt{\prod_{\alpha =1}^{g+1} \frac{\lambda -
r^+_\alpha}{\lambda - r^-_\alpha}} = Y^{-1}(\lambda) \prod_{\alpha =1}^{g+1}
(\lambda - r^+_\alpha)
\ee
In terms of these functions, the Szeg\"o kernel is
equal to \cite{Fay,calc}
\be \Psi_e(\lambda,\mu) = \frac{U_e(\lambda) +
U_e(\mu)}{2\sqrt{U_e(\lambda)U_e(\mu)}} \frac{\sqrt{d\lambda d\mu}}{\lambda -
\mu}
\label{Sz}
\ee
The squares of the Szeg\"o kernel and the bi-differential $W(\xi,\zeta)$
are always related by the identity (see Proposition 2.12 in the Fay's
book in \cite{Fay}, \cite{calc} and section \ref{A} below):
\be
\Psi_e(\xi,\zeta)\Psi_{-e}(\xi,\zeta) =  W(\xi,\zeta) +
d\omega_i(\xi)d\omega_j(\zeta)
\frac{\partial^2}{\partial z_i\partial z_j}
\log \theta_e(\vec 0|{\cal T})
\label{relation}
\ee
This allows one to express $W(\xi,\zeta)$ in (\ref{secderF}) through the
square of the Szeg\"o kernel (note that, for the half-integer characteristics,
$-e$ is equivalent to $e$).

A crucial point is that, for a peculiar even theta-characteristic $e=E$,
the Szeg\"o kernel for the Toda-chain curves (\ref{hell})
is especially simple and contributions to (\ref{secderF})
are easily evaluated.
Existence of distinguished characteristic follows from the special
shape of $Y^2 = P^2-4\Lambda^{2N} = (P-2\Lambda^{N})
(P+2\Lambda^{N})$, which implies a
distinguished separation of all the ramification points (zeroes of $Y^2$)
into two equal sets consisting of the zeroes of
$P(\lambda)\pm 2\Lambda^{N} =
\prod_{\alpha=1}^{N}(\lambda - r^\pm_\alpha)$.
Associated even theta-characteristic is denoted
through $E$ in (\ref{2der}) and later on.
Eq.(\ref{Sz}) allows one to evaluate the square of the corresponding
Szeg\"o kernel:
\be
\Psi_E^2(\lambda,\mu) = \frac{P(\lambda)P(\mu) - 4\Lambda^{2N}
+ Y(\lambda)Y(\mu)}{2Y(\lambda)Y(\mu)}
\frac{d\lambda d\mu}{(\lambda - \mu)^2}
\ee
Expanding this expression near $\mu = \infty_\pm$, one gets:
\be
\Psi_E^2(\lambda,\mu) =
\sum_{n\geq 1} \hat \Psi^2_E(\lambda)
\frac{n\lambda^{n-1}d\mu}{\mu^{n+1}}
\left(1 + O\left(P^{-1}(\mu)\right)\right)
\label{expan}
\ee
where
\be
\hat \Psi^{2\pm}_E(\lambda) \equiv
\frac{P(\lambda) \pm Y(\lambda)}{2Y(\lambda)}d\lambda =
\left\{\begin{array}{rc}
\left(1 + O(\lambda^{-2N})\right) d\lambda & {\rm near}\ \infty_\pm \\
O(\lambda^{-2N})d\lambda\ & {\rm near} \ \infty_\mp
\end{array}\right.
\label{Szk}
\ee

In order to make use of (\ref{relation}), one needs a similar
expansion for $d\omega_j(\mu)$, which is provided by
(\ref{doexp}):
\be
d\omega_j(\mu) = \sum_{n\geq 1}\frac{nd\mu}{\mu^{n+1}}
\left(\frac{1}{n}\frac{\partial h_{n+1}}{\partial a^j}\right)
\ee

Now, writing down (\ref{Wexp}) in the hyperelliptic
parametrization (with $\zeta = 1/\mu$)
and substituting (\ref{Wexp}), (\ref{expan})
and (\ref{doexp}) into (\ref{relation}), one gets
for $1 \leq n < 2N$:
\be
d\tilde\Omega^\pm_n(\lambda)  =
\lambda^{n-1}\hat \Psi^{2\pm}_E(\lambda) -
\rho^i_n d\omega_i(\lambda) \nn \\
d\tilde\Omega^n(\lambda)  =
\lambda^{n-1}(\hat\Psi^{2+}_E(\lambda) + \hat\Psi^{2-}_E(\lambda)) -
2\rho^i_n d\omega_i(\lambda)
\label{relsim}
\ee
where
\be
\rho^i_n \equiv \frac{1}{n}\frac{\partial h_{n+1}}{\partial a^j}
\partial^2_{ij}\log\theta_E(\vec 0|{\cal T})
\label{rhodef}
\ee
One can see that condition (\ref{canbas}) is fulfilled what
{\it a posteriori} justifies
the use of expansion (\ref{Wexp}) in the hyperelliptic coordinates on
the Toda-chain curves (\ref{hell}).

\subsubsection{Final step}

Now we are ready to finish evaluation of (\ref{profor}):
for $m,n = 1,\ldots,N-1$
\be
\frac{\partial^2{\cal F}}{\partial T^m\partial T^n} =
\frac{N^2}{i\pi mn} \sum_{l=1}^m lp_{ml}^{(N)}
{\res}_\infty w^{n/N} d\tilde\Omega^+_l =
\frac{N^2}{i\pi mn} \sum_{l=1}^m lp_{ml}^{(N)}
{\res}_\infty w^{n/N}
\left(\lambda^{l-1}d\lambda - 2\rho^i_l d\omega_i(\lambda)\right)
\label{FTTpro}
\ee
The first term at the r.h.s. is equal to:
\be
\sum_{l=1}^m lp_{ml}^{(N)}
{\res}_\infty w^{n/N} \lambda^{l-1}d\lambda =
{\res}_\infty
\left(w^{n/N}d P^{m/N}_+(\lambda)\right) =
{\res}_\infty
\left(P^{n/N}(\lambda)d P^{m/N}_+(\lambda)\right)
\label{FTTf1}
\ee
The second term is evaluated with the help
of (\ref{cadiresfin}) and (\ref{pvarh}):
\be
-2\left(\sum_{l=1}^m lp_{ml}^{(N)}\rho^i_l\right)
{\res}_\infty w^{n/N} d\omega_i(\lambda) =
-2\frac{\partial{\cal H}_{m+1}}{\partial a^j}
\partial^2_{ij}\log\theta_E(\vec 0|{\cal T})
\frac{\partial{\cal H}_{n+1}}{\partial a^i}
\label{FTTf2}
\ee

The sum of expressions (\ref{FTTf1}) and (\ref{FTTf2}) is:
\be
\frac{\partial^2{\cal F}}{\partial T^m\partial T^n} =
\frac{N^2}{i\pi mn}\left(
{\res}_\infty
\left(P^{n/N}(\lambda)d P^{m/N}_+(\lambda)\right)
- 2\frac{\partial{\cal H}_{m+1}}{\partial a^i}
\frac{\partial{\cal H}_{n+1}}{\partial a^j}
\partial^2_{ij}\log\theta_E(\vec 0|{\cal T})
\right)
\label{FTTf}
\ee
This completes the proof of eq.(\ref{2der}).

\subsection{Second $\alpha$-derivatives}

The remaining second derivative (needed, for example,
to complete the expression (\ref{hom}))
is just the period matrix of the
spectral curve (\ref{hell}):
\be
\frac{\partial^2{\cal F}}{\partial
\alpha^i\partial \alpha^j} = \oint_{B^i}d\omega_j = {\cal T}_{ij}
\ee

\section{Derivatives over $\Lambda$ from the SW Anzatz
\label{dirder}}
\setcounter{equation}{0}

In this section we turn to the direct proof of eq.(\ref{lns}).
Within the approach of ref.\cite{GKMMM}, eq.(\ref{lns})
is a particular case of (\ref{2der}), being the only case which
has meaning within the SW anzatz {\it per se}, without any
extension to the whole Whitham hierarchy. Thus, it deserves
a special consideration, which does {\it not} rely upon the
prepotential theory of ss.\ref{prepth} and \ref{specifics}.

Instead, we derive (\ref{lns}) from the relation (\ref{dudll}).
For this purpose, we need to know $A$-periods of the differential
$\frac{P+Y}{Y}d\lambda = 2\frac{wd\lambda}{Y}$.
According to (\ref{expan}), this is nothing but twice the square of
the Szeg\"o kernel $2\Psi^2_E(\lambda,\infty)$
for the particular even theta-characteristic $E$ and at
$\mu = \infty$.
\footnote{
In general, 1-differential with a double pole has $2g$ zeroes
and is defined modulo linear combination of $g$ holomorphic
1-differentials. This ambiguity allows one to impose
constraints on locations of any $g$ of the $2g$ zeroes.
For example, one can require the differential to have $g$
{\it double} zeroes. Such a differential is a square
of a $1/2$-differential (which can change sign when carried
along a non-contractable contour) and can be written
explicitly as a square
of the Szeg\"o kernel $\Psi^2_e(\lambda,\mu)$, where $\mu$ is the
location of the double pole. However, on peculiar curves
there can be special choices of points $\mu$ and characteristics $e$,
where the degeneracy of zeroes increases. In particular,
$\frac{wd\lambda}{Y}$ on the curve
(\ref{hell}), which has double pole at $\lambda=\infty$ on one
sheet of the curve, has a single $2g = 2N-2$-fold zero at
$\lambda = \infty$ on the other sheet. It means that, for some
theta-characteristic, all the zeroes of $\Psi^2_e(\lambda,\infty)$
should coincide. This is indeed the case
as explicitly demonstrated in the main body of the text.}

The $A$-periods of $\Psi^2$ are easily evaluated with the help of
(\ref{relation}), since $W$ has vanishing $A$-periods.
Therefore, one gets from (\ref{dudll}), (\ref{relsim})
and (\ref{rhodef}):
\be
-\frac{\partial u_k}{\partial \log\Lambda}
\frac{\partial a^i}{\partial u_k}
= 2N\Lambda^{N}\oint_{A_i} \frac{wd\lambda}{Y} =
N\oint_{A_i}{P+Y\over Y}d\lambda=\nn\\=
2N\oint_{A_i} \hat\Psi_E^2(\lambda) =
2N \rho^i_1 =
2N \frac{\partial h_{2}}{\partial a^j}
\partial^2_{ij}\log\theta_E(\vec 0|{\cal T})
\ee
(Instead of using (\ref{relsim}) and (\ref{rhodef}), one
can directly apply (\ref{relation}) together with
(\ref{candif}) and take into account that
the only differential $dv^k(\mu) = \frac{\mu^{N-k}d\mu}{Y(\mu)} =
\frac{d\mu}{\mu^k}(1 + O(\mu^{-1}))$ which does not vanish at
infinity is $dv^2$ so that
$d\hat\omega_j(\infty) = \sigma^{-1}_{jk}d\hat{v}^k(\infty) =
\sigma^{-1}_{j,2} = \partial u_2/\partial a^j$.)

This proves the relation (\ref{lns}):
\be
\frac{\partial u_k}{\partial \log\Lambda} = -2N
\frac{\partial u_k}{\partial a^i} \frac{\partial u_2}{\partial a^j}
\partial^2_{ij} \log\theta_E(\vec 0|{\cal T})
\label{lns1}
\ee
Here one can substitute $u_k$ at the both sides of the equation
by any function of $u_k$ (but not of any other arguments!), for
instance, by $h_k$ or ${\cal H}_{k+1}$. Also, $u_2 = h_2$.

In s.\ref{B} below, we give an even more explicit proof of
(\ref{lns1}) for $N=2$, i.e. for the elliptic spectral curves in the
Toda-chain parametrization (\ref{hell}),
which does not rely upon transcendental relations like (\ref{relation}).
The only transcendental ingredient of that proof will be
the well known Picard-Fuchs equation.

\section{On the relation between $\log\Lambda$-derivatives of
the Hamiltonians and the prepotential $T$-derivatives
\label{cocheck}}
\setcounter{equation}{0}
\subsection{$\log\Lambda$- versus $\log T_1$-derivatives}

From the relation (\ref{varhell})
\be
P'\delta\lambda  - \sum_k\lambda^{N-k}\delta u_k =
NP\delta\log\Lambda
\ee
it follows that
\be
\delta a^i = \oint_{A_i} \delta\lambda\frac{dw}{w} = \sum_k
\delta u_k \oint_{A_i} \frac{\lambda^{N-k}}{P'}\frac{dw}{w} +
N\delta\log\Lambda \oint_{A_i} \frac{P}{P'}\frac{dw}{w}
\ee
and
\be
\sum_k \oint_{A_i} dv^k
\left(\left.\frac{\partial u_k}{\partial\log\Lambda}
\right|_{a = const}\right)
= -N \oint_{A_i} \frac{P}{P'}\frac{dw}{w}
= -N \oint_{A_i} \frac{Pd\lambda}{Y}
\label{Lder}
\ee
This is the expression that we used in s.\ref{dirder}.

On the other hand, for $\alpha^i = T_1 a^i + O(T_2, T_3, \ldots)$
\be
\delta\alpha^i  = \alpha^i \delta\log T_1 +
T_1\oint_{A_i} \delta\lambda\frac{dw}{w} + O(T_2,T_3,\ldots)
\ee
and, for constant $\Lambda$ and $T_n = \delta_{n,1}$,
\be
\sum_k \oint_{A_i} dv^k
\left(\left.\frac{\partial u_k}{\partial\log T_1}
\right|_{\alpha = const}\right) = -\frac{\alpha^i}{T_1} =
-\oint_{A_i} \frac{\lambda dP}{Y}
\label{Tder}
\ee
Since $\lambda dP = NPd\lambda + \sum_k ku_k\lambda^{N-k}d\lambda$,
from (\ref{Lder}) and (\ref{Tder}) it follows that
\be
\left.\frac{\partial u_k}{\partial\log T_1}
\right|_{\alpha = const}
= \left.\frac{\partial u_k}{\partial\log\Lambda}
\right|_{a = const}  - ku_k = -a^i\frac{\partial u_k}{\partial a^i}
\label{basic}
\ee
The last equality follows from (\ref{dimrel}).

Relation (\ref{basic}) is, of course, true for any {\it homogeneous}
algebraic combinations of $u_k$, in particular, for $h_k$ and
${\cal H}_k$.

\subsection{Consistency check}

To check the consistency between (\ref{2der}) and (\ref{lns}),
let us take the relation (\ref{1der}),
\be
\frac{\beta}{2\pi i}{\cal H}_{k+1} = \frac{k}{T_1}
\frac{\partial{\cal F}}{\partial T_k}
\label{coch0}
\ee
and differentiate it w.r.t. $T_1$.

Then, expressing the l.h.s. with
the help of (\ref{basic}) through $\log\Lambda$-derivative of
${\cal H}_{k+1}$ and making use of (\ref{lns}) one gets:
\be
\frac{\beta}{2\pi i}\frac{\partial{\cal H}_{k+1}}{\partial\log T_1} =
\frac{\beta}{2\pi i}\left(
\frac{\partial{\cal H}_{k+1}}{\partial\log\Lambda} -
(k+1){\cal H}_{k+1}\right) =
-\frac{\beta^2}{2\pi i}
\frac{\partial{\cal H}_{k+1}}{\partial a^i}\rho^i_1
-(k+1)\frac{\beta}{2\pi i}{\cal H}_{k+1}
\label{rhs1}
\ee
On the other hand, express the r.h.s. through the second
prepotential derivative and apply (\ref{2der}):
\be
\frac{\beta}{2\pi i}\frac{\partial{\cal H}_{k+1}}{\partial\log T_1} =
k\frac{\partial^2{\cal F}}{\partial T_1\partial T_k}
-\frac{\beta}{2\pi i}{\cal H}_{k+1}  = \nn \\ =
\frac{\beta^2}{2\pi i}
\left(\frac{1}{2} {\res}_\infty
\left(P^{k/N}(\lambda)d P^{1/N}_+(\lambda)\right) -
\frac{\partial{\cal H}_{k+1}}{\partial a^i}\rho^i_1\right)
-\frac{\beta}{2\pi i}{\cal H}_{k+1}
\label{rhs2}
\ee
Since $P^{1/N}_+(\lambda) = \lambda$, one can use (\ref{Hdef}):
\be
\beta{\res}_\infty
\left(P^{k/N}(\lambda)d P^{1/N}_+(\lambda)\right)
= -2k{\cal H}_{k+1}
\ee
(since $\beta = 2N$),
and one can see that the r.h.s. of (\ref{rhs1}) and (\ref{rhs2})
are identical. This proves the consistency of
(\ref{2der}) and (\ref{lns}), or, in other words,
provides a direct derivation of (\ref{lns})
from (\ref{1der}) and (\ref{2der}). Eq.(\ref{lns})
is the particular case of (\ref{2der}) when
$T_2 = T_3 = \ldots = 0$.

\subsection{Consistency check for non-vanishing $T_2, T_3, \ldots$}

The generalization of relation (\ref{dimrel}) for the case
of generic non-vanishing time-variables can be obtained by
the substitution of eqs.(\ref{1der})-(\ref{2der}) into the
homogeneity relation (\ref{hom}). Eq.(\ref{hom}) implies that
\be
\frac{\partial{\cal F}}{\partial T^n} =
\sum_i \alpha^i
\frac{\partial^2{\cal F}}{\partial \alpha^i\partial T^n}
+ \sum_m T_m
\frac{\partial^2{\cal F}}{\partial T^m\partial T^n}
\ee
and substitution of (\ref{1der})-(\ref{2der}) gives for
$n = 1,\ldots,N_c-1$ and $T_{m\geq N}=0$
\be
\alpha^i\frac{\partial{\cal H}_{n+1}}{\partial a^i} =
\sum_m T_m\left( (m+n){\cal H}_{m+1,n+1}
+ \frac{\beta}{m}\frac{\partial{\cal H}_{m+1}}{\partial a^i}
\frac{\partial{\cal H}_{n+1}}{\partial a^j}
\partial^2_{ij}\log\theta_E(\vec 0|{\cal T})\right)
\label{bagen}
\ee
For $T_m=\delta_{m,1}$ and with the help of (\ref{lns}),
eq.(\ref{bagen}) turns into (\ref{dimrel}).

\section{Conclusion}

This paper contains a technical proof of identities
(\ref{2der}) and (\ref{lns}). The main motivation for
these by-now standard \cite{Fay,calc} calculations is to demonstrate
once again the relevance of the prepotential theory (the
quasiclassical integrability) to the study of effective actions,
beyond the Generalized Kontsevich Model (GKM) and even
the SW anzatz. This is an old claim \cite{UFN,GKMMM}, but, for any check
of such claims, one needs
at least some relatively independent results about effective actions
derived by relatively alternative methods. This time such a
result is provided, in the form of eq.(\ref{lns}), by ref.\cite{LNS}.

In this paper, we have concentrated on the simplest example of the SW
theory for the pure gauge models in $4d$ with the gauge group $SU(N)$.
One could further:

-- Evaluate higher derivatives of the prepotential. This time there
are no results from the Donaldson theory yet -- to compare with.
Still one can evaluate, at least, $\log\Lambda$-derivatives like it
is done in s.\ref{dirder} above with the help of the SW anzatz itself.

-- Consider $5d$ models, other gauge groups and include matter fields.
Considerations of s.\ref{dirder} are {\it literally} applicable (and
(\ref{lns}) remains just the same, only $\beta =2N - N_f$)
for the case
when $N_f$ is even and all the masses are pairwise coincident,
$Q(\lambda) = \Lambda^{2N-N_f} \prod_{\iota = 1}^{N_f}
(\lambda - m_\iota) = q^2$
so that $Y^2 = P^2 - 4Q = (P-2q)(P+2q)$ factorizes into two polynomials
of degree $N$. The case of different masses is more sophisticated
(this is the well-known phenomenon: the SW anzatz is much more elegant
when masses are pairwise coincident,
compare for example with ref.\cite{SCh}).

-- Analyze the UV-finite models, e.g. generic Hitchin families
of spectral curves (with non-trivial, at least elliptic bare curves)
instead of the simplest Toda-chain ones. Here one can make
contact with the recent results of refs.\cite{NY} and \cite{LA}.

It is also interesting to put the relation of the
Donaldson and SW theories to the
calculus on Riemann surfaces (i.e. to the free-field
formalism in conformal field theories) on a more solid ground.

However, the main remaining problem in the context of the present paper
is to work out the exact relation between the generating function
(\ref{z}) and the prepotential, which should extend the result
\cite{t+T} for GKM to less trivial theories.
As an intermediate step, one can consider partition function of
integrable system in Hitchin formalism, where it is
naturally expressed in form of the ($1d$) Kontsevich-like
integral.

\section*{Acknowledgments}

We acknowledge discussions with J.Ambjorn, S.Kharchev, I.Krichever, A.Losev
and N.Ne\-k\-ra\-sov.

This work was partly supported by the RFBR grants
96-02-16117 (A.Mar.) and 96-02-19085 (A.Mir.), CRDF grant RP2-132 (A.Gor),
INTAS grants 96-482 (A.Gor.), 96-518 (A.Mar.) and 93-1038 (A.Mir.)
and the Russian President's grant 96-15-96939 (A.Mor.)

\app{$A$-periods of squared Szeg\"o kernel
from the Wick theorem   \label{A}}

For the sake of completeness, we present here a derivation of
the relation (\ref{relation}). It can be found, for example,
in the Fay's book \cite{Fay}.

As most results in the calculus on Riemann surfaces, it can
be derived from the basic relation: the Wick theorem for
the correlator of fermions ($1/2$-differentials),
\be
\langle \prod_{i=1}^n \psi(\xi_i) \tilde\psi(\zeta_i) \rangle_e =
\det_{n\times n}  \langle \psi(\xi_i) \tilde\psi(\zeta_j) \rangle_e
\label{Wick}
\ee
where the l.h.s. is equal to
\be
\langle \prod_{i=1}^n \psi(\xi_i) \tilde\psi(\zeta_i) \rangle_e
= \frac{\theta_e(\sum_i\vec\xi_i - \sum_i\vec\zeta_i)}{\theta_e(\vec 0)}
\frac{\prod_{i<j} E(\xi_i,\xi_j) E(\zeta_i,\zeta_j)}
{\prod_{i,j} E(\xi_i,\zeta_j)}
\label{fercor}
\ee
and pairwise correlators at the r.h.s. are Szeg\"o kernels
(\ref{Sz}) (the particular case of (\ref{fercor}) for $n=1$
gives (\ref{Sz}) itself).
The Wick theorem is the property of the
theories with quadratic action (it is especially simple for
$1/2$-differentials that generically do not possess zero modes), and it is
valid for any characteristic (abelian gauge field) $e$, not obligatory
half-integer. All non-trivial information about the special
functions on Riemann surfaces (complex curves) one usually needs is
effectively summarized in (\ref{Wick}) and (\ref{fercor}).

In particular, in order to derive (\ref{relation}) one takes
the 4-point correlator ($n=2$) and considers the limit where
$\xi_1 = \xi_2$ and $\zeta_1 = \zeta_2$. The only tricky thing is that
one simultaneously adjusts $e$ (thus, this is a kind of the
"double-scaling limit").

First, write down (\ref{Wick}) explicitly for the case of $n=2$:
\be
\theta_e(\vec 0)\theta_e(\vec\xi_1+\vec\xi_2 - \vec\zeta_1-\vec\zeta_2)
E(\xi_1,\xi_2)E(\zeta_1,\zeta_2) = \nn \\ =
\theta_e(\vec \xi_1-\vec\zeta_1)\theta_e(\vec\xi_2 - \vec\zeta_2)
E(\xi_1,\zeta_1)E(\xi_2,\zeta_2)  -
\theta_e(\vec \xi_2-\vec\zeta_1)\theta_e(\vec\xi_1 - \vec\zeta_2)
E(\xi_2,\zeta_1)E(\xi_1,\zeta_2)
\label{prom2}
\ee
Now, put $\xi_1=\xi_2=\xi$. Then (\ref{prom2}) becomes
\be
\frac{\theta_e(2\vec\xi - \vec\zeta_1 - \vec\zeta_2)\theta_e(\vec 0)
E(\zeta_1,\zeta_2)}
{\theta_e(\vec\xi - \vec\zeta_1)\theta_e(\vec\xi - \vec\zeta_2)
E(\xi,\zeta_1)E(\xi,\zeta_2)} =
\partial_\xi \log\left(
\frac{\theta_e(\vec\xi - \vec\zeta_1) E(\xi,\zeta_2)}
{\theta_e(\vec\xi - \vec\zeta_2) E(\xi,\zeta_1)}
\right) = \nn \\ =
\partial_\xi \log\frac{E(\xi,\zeta_2)}{E(\xi,\zeta_1)}
+ d\omega_i(\xi)\left(\frac{\partial\log\theta_e(\vec\xi - \vec\zeta_1)}
{\partial\xi^i} - \frac{\partial\log\theta_e(\vec\xi - \vec\zeta_2)}
{\partial\xi^i}\right)
\label{prom}
\ee
Next, let us use the fact that
$\theta_e(\vec\xi) = \theta_*(\vec\xi + \vec e)$ and substitute
$\vec e$ by $\vec e - (\vec\xi - \vec\zeta_2)$ to get from (\ref{prom}):
\be
\frac{\theta_e(\vec\xi - \vec\zeta_1)\theta_{-e}(\vec\xi - \vec\zeta_2)}
{\theta_e(\vec 0)\theta_{-e}(\vec\zeta_1 - \vec\zeta_2)}
\frac{E(\zeta_1,\zeta_2)}{E(\xi,\zeta_1)E(\xi,\zeta_2)} =  \nn \\ =
\partial_\xi \log\frac{E(\xi,\zeta_2)}{E(\xi,\zeta_1)}
+ d\omega_i(\xi)\left(\frac{\partial\log\theta_e(\vec\zeta_2 - \vec\zeta_1)}
{\partial\xi^i} - \frac{\partial\log\theta_e(\vec 0)}
{\partial\xi^i}\right)
\ee
Finally, put $\zeta_1=\zeta_2=\zeta$. Then the last relation turns
into (\ref{relation}):
\be
\Psi_e(\xi,\zeta)\Psi_{-e}(\xi,\zeta) =
\frac{\theta_e(\vec\xi - \vec\zeta)\theta_{-e}(\vec\xi - \vec\zeta)}
{\theta_e(\vec 0)\theta_{-e}(\vec 0) E^(\xi,\zeta)} =
\partial_\xi\partial_\zeta \log E(\xi,\zeta)
+ d\omega_i(\xi)d\omega_j(\zeta)
\left(\frac{\partial^2\log\theta_e(\vec 0)}
{\partial\xi^i\partial\xi^j}\right)
\ee

\app{Explicit derivation in elliptic case \label{B}}
\sapp{Calculation based on Thomae formula and
Picard-Fuchs equations}

The SW curve:
\be\label{c}
w + \frac{1}{w} = P(\lambda), \nn\\
P(\lambda) = \lambda^{N}-\sum_{k=2}^{N} u_k\lambda^{N-k}, \nn\\
Y^2(\lambda) = P^2 - 4
\ee

Periods:
\be
a^i = \oint_{A_i} dS = \int_{A_i} \lambda \frac{dw}{w} =
\oint_{A_i} \frac{\lambda P' d\lambda}{\sqrt{P^2-4}}, \nn \\
\sigma^{ik} = \frac{\partial a^i}{\partial u_k}  =
\oint_{A_i} dv^k = \oint_{A_i} \frac{\lambda^{N-k}d\lambda}{\sqrt{P^2-4}},
\nn \\
i = 1,\ldots,N-1, \ \ \ k = 2,\ldots,N
\ee

For the direct check of eq.(\ref{lns}), one needs to use
integration by parts,
\be
d\left(\frac{\lambda^{k+1}}{\sqrt{P^2-4}}\right) =
(k+1)\frac{\lambda^kd\lambda}{\sqrt{P^2-4}} -
\frac{\lambda^{k+1}PdP}{(P^2-4)^{3/2}}
\ee
and the Thomae formula for the even non-singular
theta-constant \cite{Fay,calc},
\be
\theta^4_e(\vec 0) = \left(\det \sigma^2
\right)\prod_{1\leq\alpha<\beta\leq g+1}
(r^+_\alpha - r^+_\beta)(r^-_\alpha - r^-_\beta)
\ee

In the case of genus $g=1$,  $P = \lambda^2 - h$,
$a=2\oint_{A_i}{\lambda^2 d\lambda\over\sqrt{P^2-4}}$,
$\frac{\partial a}{\partial h} = \sigma =
\oint_{A_i}{d\lambda\over\sqrt{P^2-4}}$.
Then
\be
\frac{\partial\sigma}{\partial h} =
\oint_A \frac{Pd\lambda}{(P^2-4)^{3/2}}
=-\frac{1}{2}\oint_A \frac{d\lambda}{\lambda^2\sqrt{P^2-4}}
\ee
since
\be
d\left(\frac{1}{\lambda\sqrt{P^2-4}}\right) =
-\frac{d\lambda}{\lambda^2\sqrt{P^2-4}} -
\frac{PdP}{\lambda(P^2-4)^{3/2}}
\ee
and $dP = 2\lambda d\lambda$.
Further, since
\be
d\left(\frac{\sqrt{P^2-4}}{\lambda}\right) =
-\frac{\sqrt{P^2-4}}{\lambda^2}d\lambda +
\frac{PdP}{\lambda\sqrt{P^2-4}} = \nn \\
= \frac{2(\lambda^2 - h)\lambda^2 -
\left((\lambda^2 - h)^2-4\right)}{\lambda^2\sqrt{P^2-4}}d\lambda =
\left(\lambda^2 - \frac{h^2-4}{\lambda^2}\right)
\frac{d\lambda}{\sqrt{P^2-4}}
\ee
one gets the Picard-Fuchs equation:
\be
4(4-h^2)\frac{\partial\sigma}{\partial h} = a
\ee

With the help of these relations, one can easily prove (\ref{lns})
for genus $g=1$. With the help of the Thomae formula, it can be
rewritten as follows ($\beta = 2N = 4$):
\be
{\beta\over\pi i}\left(\frac{\partial h}{\partial a}\right)^2
\frac{\partial \log\theta_E(\vec 0)}{\partial{\cal T}} =
{1\over \pi i}\left(\frac{\partial h}{\partial a}\right)^2
\frac{\partial h}{\partial{\cal T}}
\frac{\partial\log (4\sigma^2  r^+r^-)}{\partial h}
= a\frac{\partial h}{\partial a}-2h,
\label{lnsg1}
\ee
In accordance with (\ref{dimrel}), the r.h.s. is equal to
$-\partial h/\partial\log\Lambda$. Here
\be
r^\pm = \sqrt{h \pm 2}
\ee
and  \cite{Fay,calc}
\be
\frac{\partial {\cal T}}{\partial h} =
\sum_\alpha \hat\omega^2(\lambda_\alpha)
\frac{\partial\lambda_\alpha}{\partial h} =
\frac{1}{\pi i\sigma^2} \sum_\alpha \frac{1}{\hat Y^2(\lambda_\alpha)}
\frac{\partial\lambda_\alpha}{\partial h}
\ee
where the four ramification points $\lambda_\alpha = \{ \pm r^\pm \}$, i.e.
\be
\frac{\partial {\cal T}}{\partial h} =
-\frac{1}{2\pi i\sigma^2 (r^+r^-)^2} =
-\frac{1}{2\pi i\sigma^2(h^2-4)}
\ee
Finally,
\be
\frac{\partial \log(r^+r^-)}{\partial h} =
\frac{1}{2(r^+)^2} + \frac{1}{2(r^-)^2} = \frac{h}{h^2-4}
\ee
All together these relations prove (\ref{lnsg1}):
\be
{1\over\pi i}\left(\frac{\partial h}{\partial a}\right)^2
\frac{\partial h}{\partial{\cal T}}
\frac{\partial\log (4\sigma^2  r^+r^-)}{\partial h} = \nn \\ =
-\frac{2\sigma^2(h^2-4)}{\sigma^2}\left[
{a\over 2\sigma(4-h^2)}+\frac{h}{h^2-4}\right] =
{a\over\sigma}-2h= a\frac{\partial h}{\partial a}-2h
\ee

It is an interesting exercise to prove (\ref{lns})
for any $N>2$ by the method of this Appendix,
i.e. to derive it directly from the Thomae formula and
Picard-Fuchs equations.

\sapp{Calculation with the help of elliptic integrals}

In the case of genus $g=1$, one can also use the rich
collection of formulas for elliptic integrals and elliptic
functions from the standard textbooks. We use as a main source
here the books \cite{BE}. To keep closer contact with these books, we
change our definition of the $\theta$-function to the standard one.

First, we express the main integrals through the elliptic
integrals and then rewrite them via $\theta$-functions. To this end,
we need to fix integration contours. We choose the $A$-cycle to encircle
the points $\lambda=r^-$ and $\lambda=r^+$. Then,
\be
a=2\oint_{A}{\lambda^2 d\lambda\over\sqrt{P^2-4}}={2\over\pi}r^+
E(k)\\
\frac{\partial a}{\partial h} = \sigma =
\oint_{A}{d\lambda\over\sqrt{P^2-4}}={1\over\pi r^+}K(k)
\ee
where $K(k)$ and $E(k)$ are complete elliptic integrals of the first and the
second kinds respectively, the elliptic modulus $k={2\over r^+}$ so that
the complimentary modulus
$k'={r^-\over r^+}$. Complete elliptic integrals, as well as their modulus,
can be written through $\theta$-constants \cite{BE}:
\be\label{ceit}
K={\pi\over 2}\theta_3^2,\ \ \ \ k={\theta_2^2\over\theta_3^2},\ \ \ \
k'={\theta_4^2\over\theta_3^2},\ \ \ \
E={\theta_3^4+\theta_4^4\over 3\theta_3^4}K-{1\over 12K}{\theta_1'''\over
\theta_1'}
\ee
The latter expression can be reduced to
\be
E=-{2i\over\theta_3^2}\partial_\tau\log\theta_2
\ee
with the help of equation $\theta_a''=4\pi i\partial_\tau\theta_a$,
the Jacobi identities
\be
\theta_1'=\pi\theta_2\theta_3\theta_4,\ \ \ \
\theta_2^4+\theta_4^4=\theta_3^4
\ee
and the evident derivatives
\be
\theta_2^4={4 i\over\pi}\partial_\tau\log{\theta_4\over\theta_3},\ \ \ \
\theta_3^4={4 i\over\pi}\partial_\tau\log{\theta_4\over\theta_2},\ \ \ \
\theta_3^4\partial_\tau\log{\theta_4\over\theta_3}=
\theta_2^4\partial_\tau\log{\theta_4\over\theta_2}
\ee
of the addition formulas
\be
\theta_1^2(v)\theta_2^2=\theta_4^2(v)\theta_3^2-\theta_3^2(v)\theta_4^2\\
\theta_1^2(v)\theta_3^2=\theta_4^4(v)\theta_2^2-\theta_2^2(v)\theta_4^2\\
\theta_1^2(v)\theta_4^2=\theta_3^2(v)\theta_2^2-\theta_2^2(v)\theta_3^2\\
\theta_4^2(v)\theta_4^2=\theta_3^2(v)\theta_3^2-\theta_2^2(v)\theta_2^2
\ee
Collecting all pieces together, one can write
\be
\left(a{\partial h\over\partial a}-2h\right)\left(
\partial a\over\partial h\right)^2=a\sigma-2h\sigma^2=
{1\over\pi i}\partial_\tau\log\left(\theta_3\theta_4\right)
\ee
To simplify this  expression further,
one can use the duplication formula \cite{BE}
\be
\theta_3(\tau)\theta_4(\tau)=\theta_4^2(2\tau)
\ee
Finally, let us note that the value of $\tau$ used in formulas
(\ref{ceit}) -- $\tau=i{K'\over K}$ \cite{BE}
is {\it half period} of the curve (\ref{c}), since
${\cal T}={\oint_{B}{d\lambda\over\sqrt{P^2-4}}\over
\oint_{A}{d\lambda\over\sqrt{P^2-4}}}=2i{K'\over K}$. This
restores the correct form of (\ref{lnsg1})
\be
a\sigma-2h\sigma^2=
{4\over\pi i}\partial_{{\cal T}}\log\theta_4
\ee
Therefore, with this choice of the integration contour, $\theta_E=\theta_4$.

\end{document}